      [1999/12/01 v1.4c Il Nuovo Cimento]
\documentclass{cimento}

\usepackage{graphicx}  
\title{UHE Leptons and Neutrons feeding Precessing $\gamma$ Jet\\
 in GRBs - SGRs:
 A  SGR 1806-20  link to EeV C.R.?}
\author{D.Fargion\from{ins:x},M.Grossi\from{ins:x}
} \instlist{\inst{ins:x} Physics Department and INFN Rome
University La
Sapienza, Rome, Italy} 

\begin{document}


\maketitle

\begin{abstract}
Soft Gamma Repeaters are widely believed to occur as isotropic
Magnetar explosion. We suggest on the contrary  that they may be
described by  thin collimated spinning and precessing gamma jets,
flashing and  blazing along the  line of sight. The jet (for SGRs)
maybe powered by an accretion disk in binary system and it
produces huge outflows and   blazing { features} oscillating mode
as observed in the light curve of  the Giant Flare from SGR
1806-20. The precessing and spinning nature {of the blazing gamma
jets} reflects, at smaller intensity, the same behaviour observed
in {short} GRBs Jetted Supernova, as well as re-brightening and
bumps in their afterglows. The SGR $\gamma$ beam maybe powered by
Inverse Compton Scattering (ICS) or by synchrotron radiation of
electron pairs  respectively at GeVs or PeVs energies. In  the
latter case, tens of PeV leptons (muons later decaying into PeV
electrons) might be originated while EeV nucleons Jets
(protons-neutrons) are in photopion equilibrium with infrared
photons surrounding the source. The neutron Jet might survive and
remain collimated. It maybe already detected as an EeV cosmic ray
anisotropy in the AGASA map pointing toward observable known SGRs
: SGR 1900 +14 and SGR 1806-20. If the SGR-EeV connection is
correct, a parasite trace of PeVs neutrinos as well PeV-TeVs gamma
rays might be found in present or future data records, among the
other CR array detectors, and inside the Amanda and Milagro
volumes.

\end{abstract}

\section{Precessing Jets in GRBs and SGRs : the blazing engine}

\begin{figure}
\includegraphics[width=1.3in]{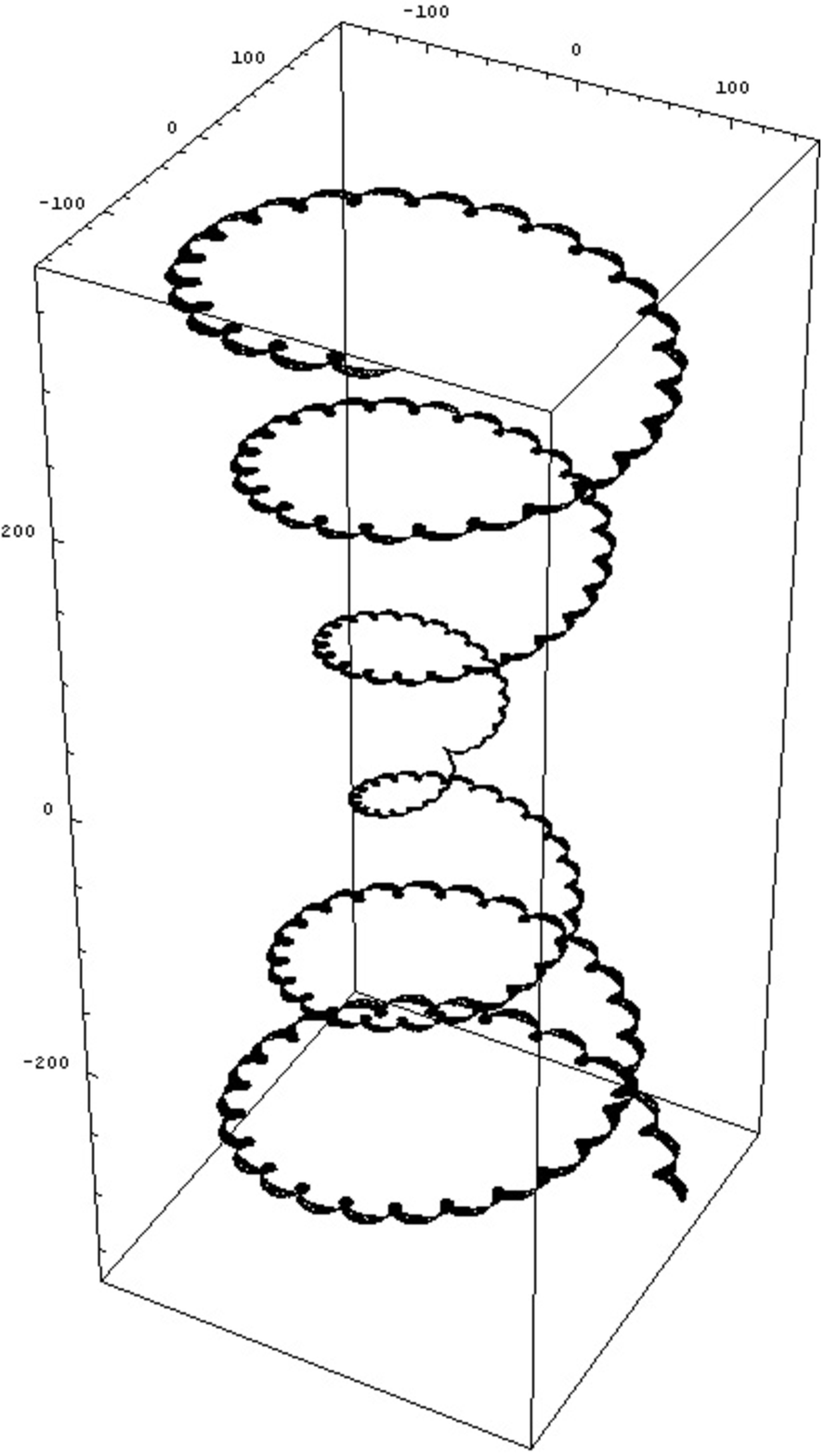}     
\includegraphics[width=1.8in]{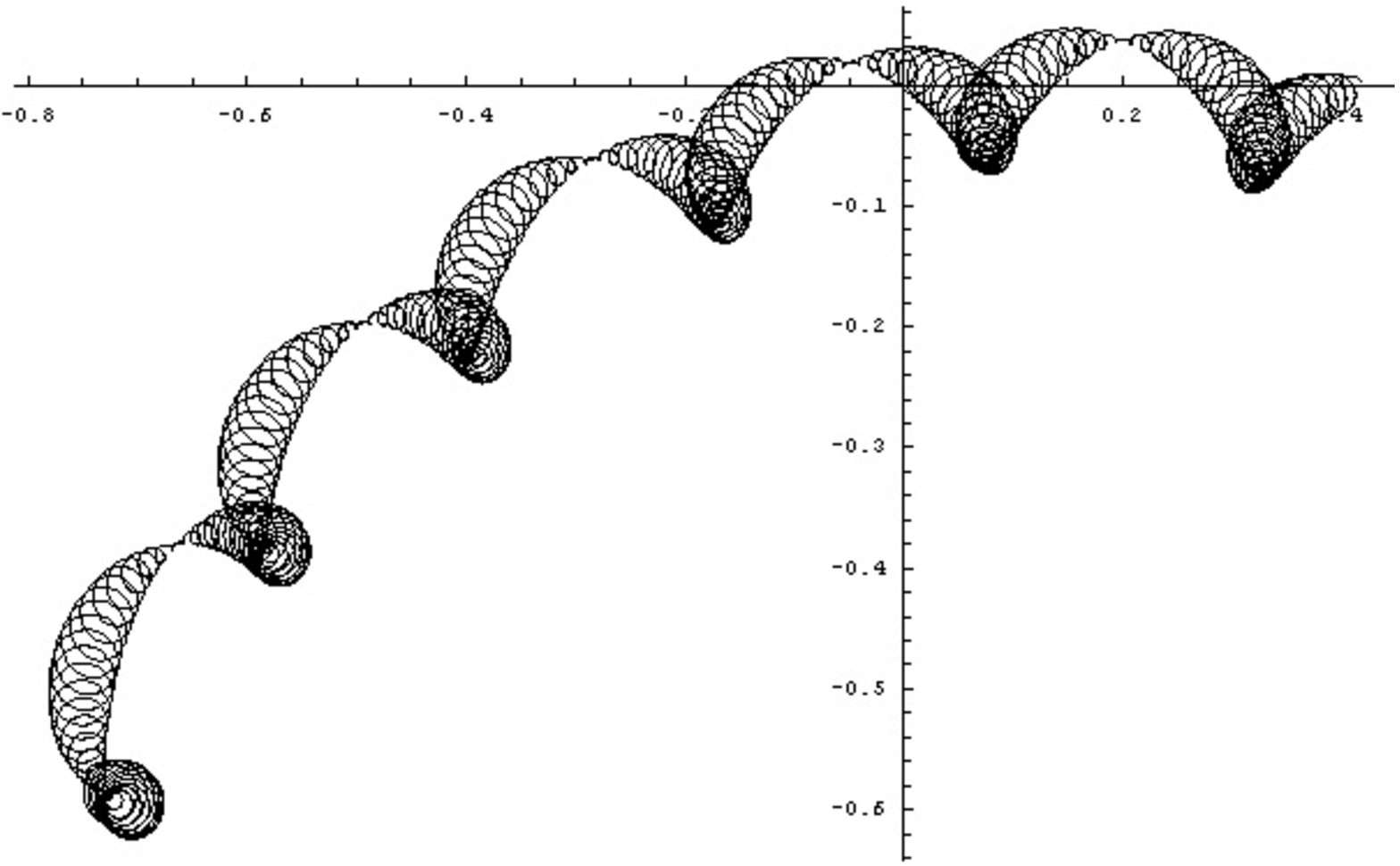}     
\includegraphics[width=1.8in]{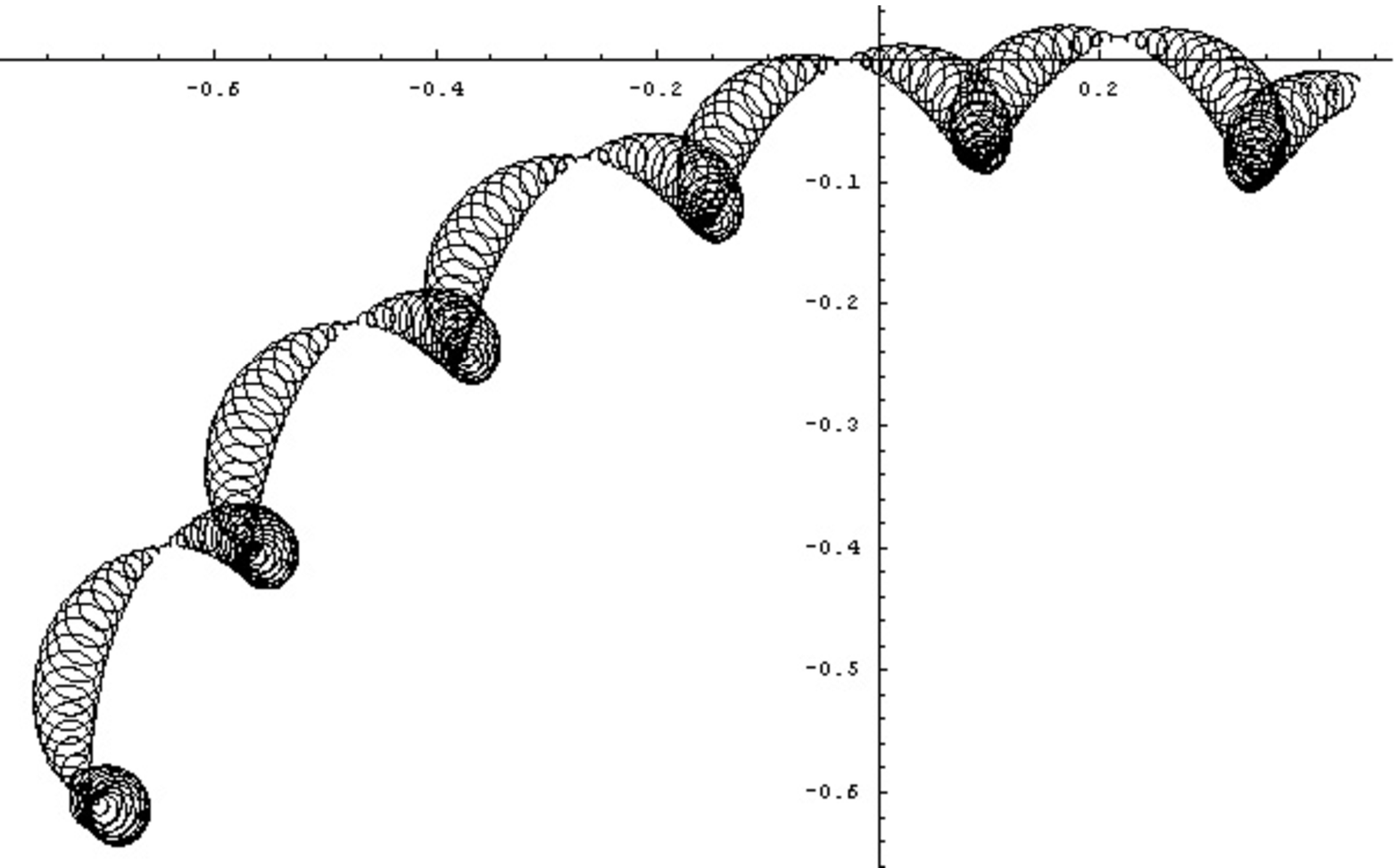}     
\caption{A possible 3D structure view of the precessing jet
obtained with non linear precessing, while spinning $\gamma$ jet;
at its centre the "explosive" SN-like event for a GRB  or a steady
binary system system for a SGRs where an accretion disc around a
compact object powers a collimated precessing jet. The Lorentz
factor is $\gamma_e = 10^9$, corresponding to a $\sim PeV$
electron pair energy; two different 2D trajectory of the
precessing jet and of the way it blazes the observer along the
line of sight  at the center, by  a fine tuned beaming to the
observer, while on the right the off-axis flashes of the same jet}
\label{spinning_GRB}
\end{figure}

\begin{figure}

\includegraphics[width=2.5in]{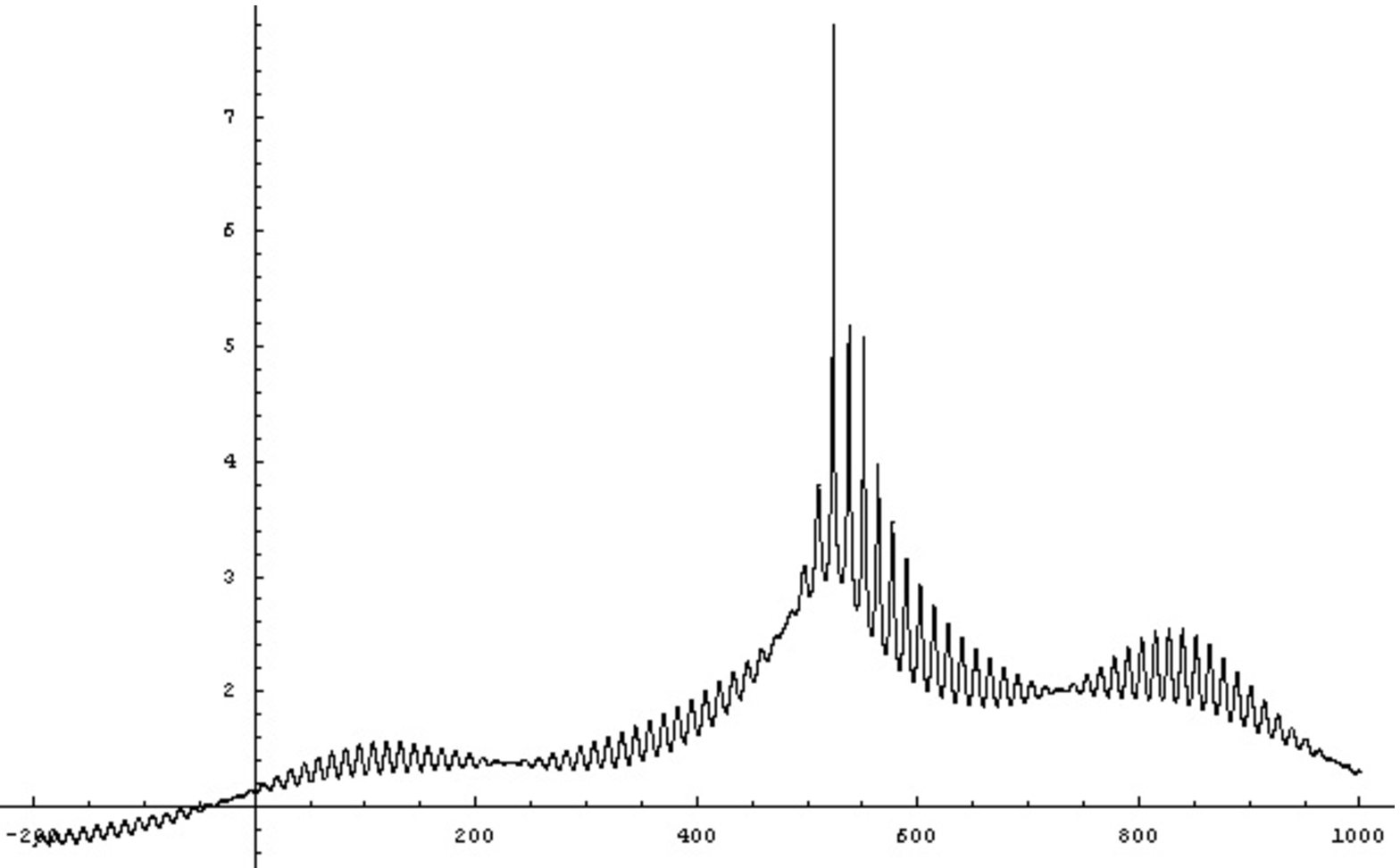}     
\includegraphics[width=2.5in]{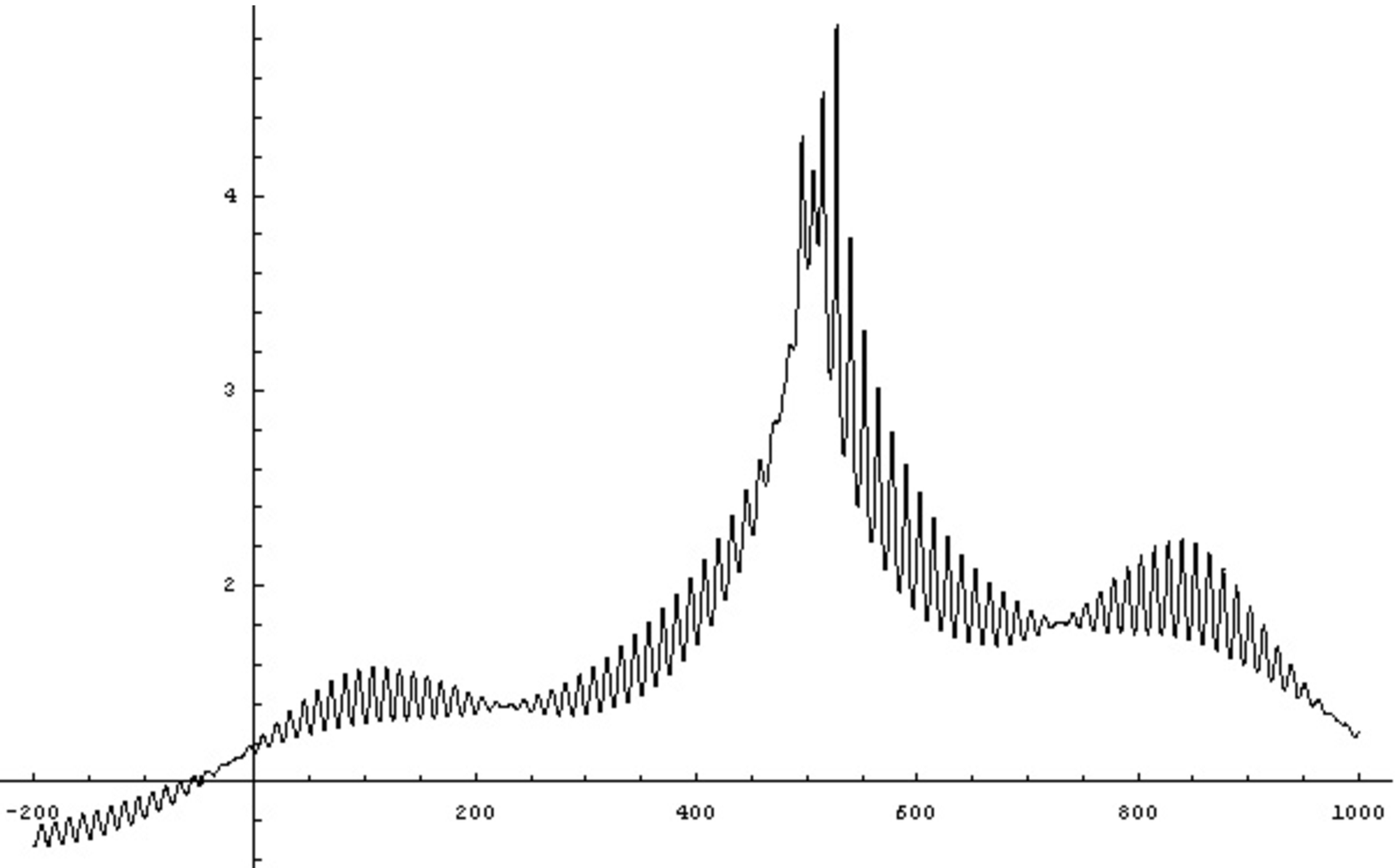}     
\caption{A close up of the two corresponding light curve profile
({ left panel; the multiple oscillatory signals may mimic  the
oscillatory bumps in SGR $\gamma$ and the huge amplification of
giant flare, while the multi-precessing tracks of the Jet may lead
to re-brightening and multi-bumps  in the light profile of the GRB
X afterglows. { Right panel}; note that the off-axis beaming
induce a different SGR profile and a much limited
amplification}).} \label{profile_SGR}
\end{figure}

The huge GRBs luminosity may be { due} to high collimated and
aligned blazing Jet powered by a Supernova output; this explains
the rare SN-GRB connection (while on the line of sight) and the
apparent GRB 990123 extraordinary power (billions of Supernova
luminosity). This also explain the rarer, because nearer, GRB-SN
event as the April 1998 one, whose Jet was off axis (increasing
its probability detection) , but whose GRB luminosity was
extremely low. This beaming selection in wider cosmic volumes
explains the puzzling evidence (the Amati-Ghisellina law) of
harder and powerful GRBs at higher and higher distances. This
"law" is  for isotropic Fireball explosion, contrary to the
opposite cosmic trend  required by the Hubble-Friedmann law: the
further the distances, the larger  the redshifts and the softer
the expected GRB event. To make GRB-SN in energy equipartition the
Jet must be very collimated $\frac{\Omega}{\Delta\Omega}\simeq
10^{8}-10^{9}$ (Fargion,Salis 1995;Fargion 1999; Fargion,Grossi
2005); because of the statistics between GRB-SN rates, its Jet
decaying  activity ($ \dot{L}\simeq (\frac{t}{t_o})^{-\alpha}$,
$\alpha \simeq 1$) must hold for long timescales, $t_o \simeq 10^4
s$. Similar issues arise from  the surprising giant flare from the
soft gamma repeater SGR 1806-20 that occurred on 2004 December 27:
if it has been radiated isotropically ({ as assumed by} the the
magnetar model), most of (if not all) the magnetic energy stored
in the neutron star (NS) should have been consumed at once. On the
contrary we think that a thin collimated precessing jet
$\dot{E}_{SGR-Jet}\simeq 10^{36}-10^{38} erg s^{-1}$, blazing
on-axis, may be the source of such apparently (the inverse of the
solid beam angle $\frac{\Omega}{\Delta\Omega}\simeq
10^{8}-10^{9}$) huge bursts $\dot{E}_{SGR-Flare}\simeq
10^{38}\cdot \frac{\Omega}{\Delta\Omega} \simeq 10^{47} erg
s^{-1}$ with a moderate (X-Pulsar, SS433) Jet output power. In our
model, the temporal evolution of the angle between the jet
direction and the rotational axis {of the  (NS)} can be expressed
as $ \theta_1 (t) = \sqrt{\theta_x^2 +  \theta_y^2}$,
where\\

 $ \theta_x(t) =  \sin (\omega_b t + \phi_{b}) + \theta_{psr}
\cdot \sin( \omega_{psr} t + \phi_{psr})\cdot Abs(\sin( \omega_N t
+ \phi_N)) + \theta_s \cdot \sin (\omega_s t+ \phi_{s}) + \theta_N
\cdot \sin (\omega_N t + \phi_N) + \theta_x(0)$  and \\

 $\theta_y(t) =  \theta_a \cdot \sin \omega_0 t  + \cos ( \omega_b t + \phi_{b})+ \theta_{psr} \cdot \cos (\omega_{psr} t +
\phi_{psr})\cdot Abs(\sin( \omega_N t + \phi_N)) + \theta_s \cdot
\cos (\omega_s t+ \phi_{s}) + \theta_N \cdot \cos (\omega_N t +
\phi_N)) + \theta_y(0)$. Where $\gamma$ is the Lorentz factor of
the relativistic particles {of the jet}.
\begin{table}
\begin{tabular}{lll}
\hline \hline
  $\gamma = 10^9$  & $\theta_a=0.2$ & $\omega_a =1.6 \cdot 10^{-8}$ rad/s\\
  $\theta_b=1$ &  $\theta_{psr}$=1.5 $\cdot 10^7$/$\gamma$ & $\theta_N$=$5 \cdot 10^7$/$\gamma$ \\
$\omega_b$=4.9 $\cdot 10^{-4}$ rad/s &  $\omega_{psr}$=0.83 rad/s
& $\omega_N $=1.38 $\cdot 10^{-2}$ rad/s \\
$\phi_{b}=2\pi - 0.44$ &$\phi_{psr}$=$\pi + \pi/4$ & $\phi_N$=3.5
$\pi/2 + \pi/3$ \\
$\phi_s \sim \phi_{psr}$ & $\theta_s$=1.5 $\cdot 10^6$/$\gamma$ & $\omega_s = 25$ rad/s \\
 \hline \hline
\end{tabular}
\caption{The parameters adopted for the jet model represented in
Fig. \ref{spinning_GRB}} \label{jet_parameters}
\end{table}

The simplest  way to produce the $\gamma$ emission  would be by
ICS of GeVs electron pairs onto thermal infra-red photons. However
also electromagnetic showering of PeV electron pairs by
synchrotron emission in galactic fields, ($e^{\pm}$ from muon
decay) maybe the progenitor of the $\gamma$ blazing jet. The same
muons showering may occur in GRB. In particular, muon bundles have
the advantage to avoid the opacity and escape the dense
GRB-SN-isotropic radiation field (Fargion,Grossi 2005). However
most of these IC PeVs showering in GRB-SN  scenario  degrades the
electrons to tens GeVs energy  leading to $\gamma$ Jet by ICS.
Here we propose {also that}  the emission of SGRs is due to a
primary hadronic jet producing ultra relativistic $e^{\pm}$ (1 -
10 PeV) from hundreds PeV pions, $\pi\rightarrow \mu \rightarrow
e$, or EeV neutron decay in flight: primary protons can be
accelerated by the large magnetic field of the NS up to EeV
energy. The protons could emit directly soft gamma rays via
synchrotron radiation with the galactic magnetic field
($E_{\gamma}^p \simeq 10 (E_p/EeV)^2 (B/2.5 \cdot 10^{-6} \, G)$
keV), but the efficiency is poor because of the too long timescale
of proton synchrotron interactions. Photopion production must
occur to produce the observed neutron excess, a process that also
gives birth to neutral and charged pions. %
The energy of the thermal photons necessary to produce pions is
$\epsilon_{\gamma} = 0.2$ GeV$^2/E_p$ = 0.2 eV for $E_p = $
10$^{18}$ eV.  Photopion production with ambient galactic IR
photons ($p + \gamma_{IR} \rightarrow \Delta^+ \rightarrow n +
\pi^+$) is favoured compared to proton-proton collisions ($p + p
\rightarrow n + p + N\pi$)(Medina-Tanco, G.A.,  \& Watson, A.A,
2001). Charged pions (born with roughly a tenth of the original
energy of the proton) decay into muons and then into electrons
with $E_e \leq 10^{16}$ eV.

By interacting with the local galactic magnetic field such
electrons lose energy via synchrotron radiation, $
E_{\gamma}^{sync} \simeq 4.2 \times 10^6 \left(\frac{E_e}{5 \cdot
10^{15} \: eV} \right)^2 \left(\frac{B}{2.5 \cdot 10^{-6} \; G}
\right) \: eV$, { with a characteristic timescale}  $ t^{sync}
\simeq 1.3 \times 10^{10} \left(\frac{E_{e}}{5 \cdot 10^{15} eV}
\right)^{-1} \left(\frac{B}{2.5 \cdot 10^{-6} \, G} \right)^{-2}
\: s $.
{ This mechanism  would produce} a few hundreds keV radiation as
it is observed in the intense $\gamma$-ray flare from SGR 1806-20.
The Larmor radius is about two orders of magnitude smaller than
the synchrotron interaction length and this may imply that the
aperture of the jet is spread by the magnetic field, $
\frac{R_L}{c} \simeq 4.1 \times 10^{8}
  \left(\frac{E_{e}}{5 \cdot 10^{15} eV} \right)
\left(\frac{B}{2.5 \cdot 10^{-6} \, G} \right)^{-1} \: s$. In
particular a thin ($\Delta \Omega \simeq 10^{-9}-10^{-10} $ $sr$)
precessing jet  from a pulsar may naturally explain the negligible
variation of the spin frequency $\nu=1/P$ after the giant flare
($\Delta \nu < 10^{-5}$ Hz). Indeed it seems quite unlucky that a
huge ($E_{Flare} \simeq 5 \cdot 10^{46} erg$) explosive event (as
the needed mini-fireball by a magnetar model (Duncan et all 1992))
is not leaving trace in the rotational energy of the SGR 1806-20,
$ E_{rot}= \frac{1}{2} I_{NS} \omega^2 \simeq 3.6 \cdot 10^{44}
\frac{P}{7.5 s}^{-2} \left( \frac{I_{NS}}{10^{45} g \, cm^2}
\right) erg \label{Erot} $. The consequent fraction of energy lost
after the flare must be severely bounded :
$\frac{\Delta(E_{Rot})}{E_{Flare}} \leq 10^{-6}$. We foresee that
if the role of nucleons  as primaries of the soft gamma emission
of SGR 1806-20 is correct, the giant flare might also be source of
a prompt (and possibly repeating) $\gamma$-induced shower (made by
$\gamma$ photons from the decay of PeV neutral pions) that may
have been detected by Milagro in correspondence with the 2004
December 27 flare. We also expect that a rich component of the EeV
neutrons (or protons from their
decay) might appear 
in the AUGER or HIRES detectors, in rough coincidence with this
event. Because of the delayed  arrival time of protons, one should
expect also a long and persistent UHECR afterglow. Finally a
signal of secondary muons at PeV energies, induced by high energy
neutrinos from the SGR, might occur in Amanda. To conclude, we
imagine  that if the precessing jet model gives a correct
interpretation  of the properties of SGRs, SGR 1806-20 will  still
be active in the next months and this year.

\begin{figure}
\includegraphics[width=2.5in]{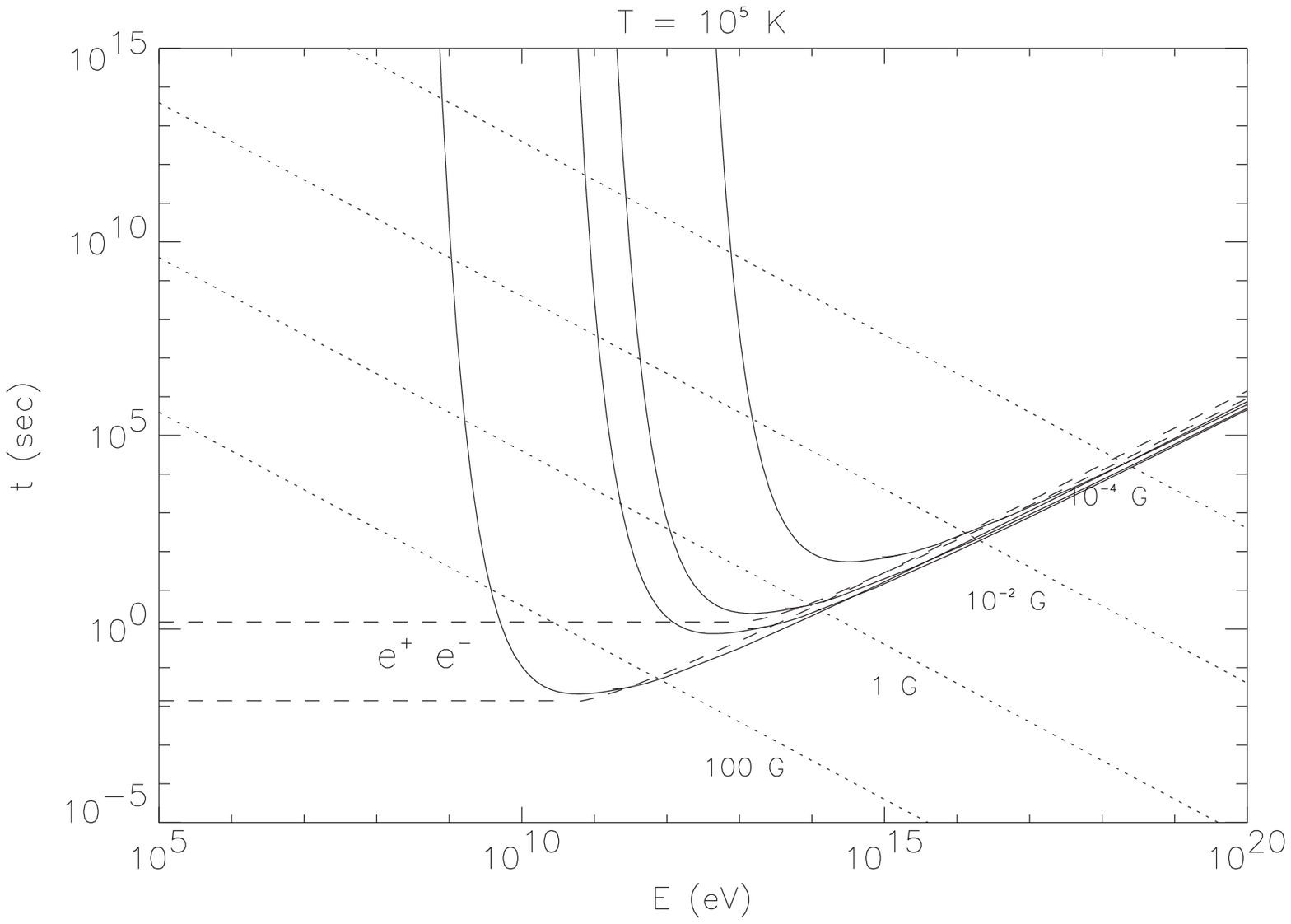}
\includegraphics[width=2.5in]{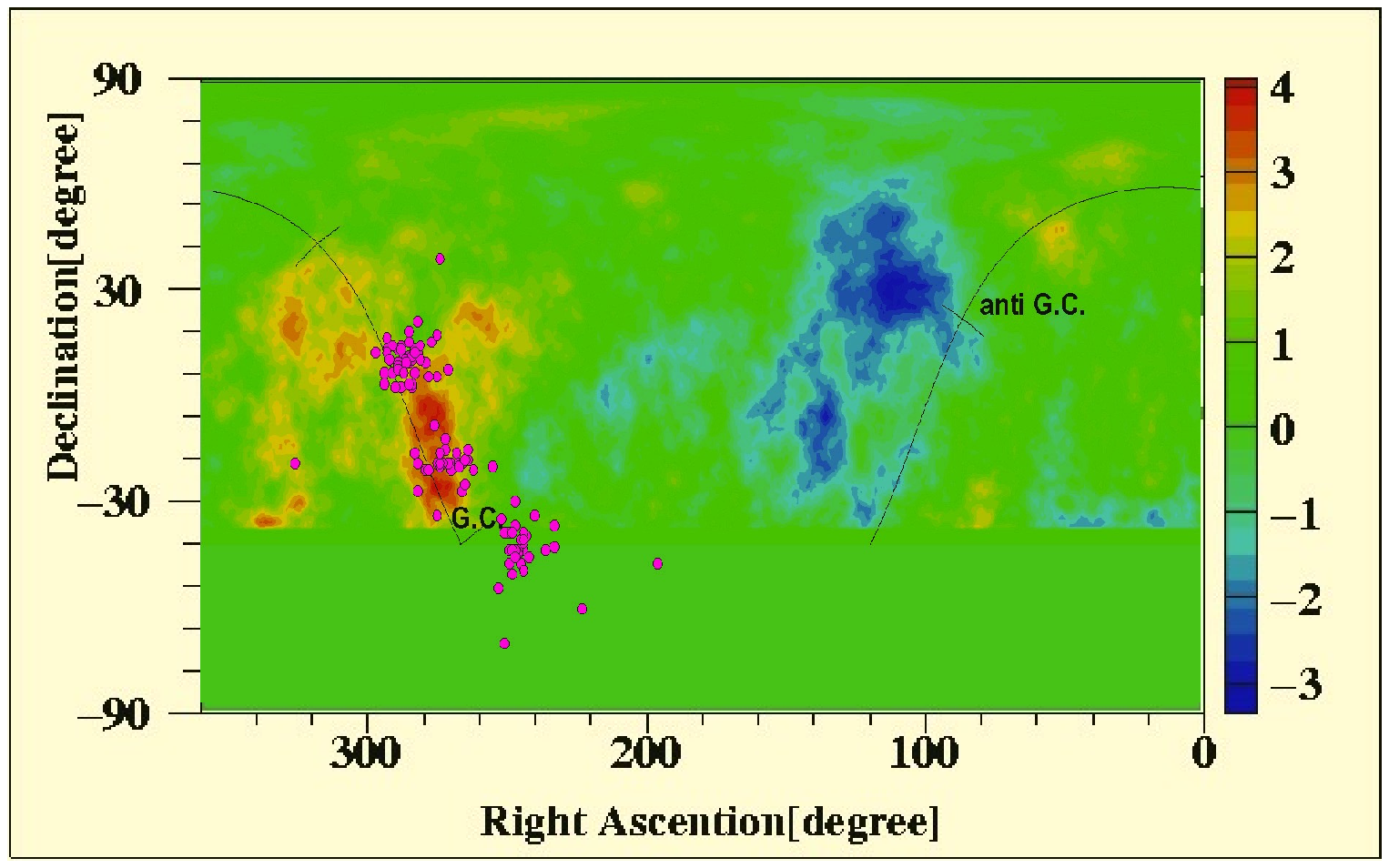}
 \caption{Left panel:
The Supernova opacity (interaction length) for PeV electrons at
different times ; PeVs muons Jets may overcome it and decay later
in $\gamma$ showering electrons (see for details  Fargion, Grossi
2005). Right Panel: The correlation between the BATSE data, the
AGASA discovery of an anisotropy in the arrival direction of EeV
CRs near the galactic center  and the position of the galactic
SGRs. The three clusters of the data from BATSE in the left-hand
side of the map correspond from top to bottom  SGR 1900+14, SGR
1806-20, SGR 1627-41, recorded during 1997-2000. AGASA was unable
to record SGR 1627-41 because below its horizons.} \label{EEV-SGR}
\end{figure}

\end{document}